\documentclass[12pt]{iopart}

\usepackage{graphicx}

\begin{document}

\article[Spatial distribution of the edge-states]{Quantum phases: 50 years of the Aharonov-Bohm Effect and 25 years of the Berry phase}{Inferring the transport properties of edge-states formed at quantum Hall based Aharonov-Bohm interferometers theoretically}

\author{E. Cicek$^1$ and A. Siddiki$^{2,3}$}

\address{$^1$ Trakya University, Department of Physics, 22030 Edirne, Turkey}
\address{$^2$Physics Department, Faculty of Arts and Science, 48170-Kotekli, Mugla, Turkey}
\address{$^3$Physics Department, Faculty of Science, Istanbul University, 34134 Vezneciler-Istanbul, Turkey}

\begin{abstract}
Here, we report on our results where self-consistent calculations are performed to investigate the interference conditions, numerically. We employ the successful 4$^{th}$ order grid technique to obtain the actual electrostatic quantities of the samples used at the quantum Hall based Aharonov-Bohm interferometers. By knowing the electron density distribution we calculate the spatial distribution of the edge-states, which are considered as mono-energetic current channels. Our results are in accord with the experimental findings concerning the electron density distribution. Finally, we also comment on the ``optimized'' sample design in which highest visibility oscillations can be measured.
\end{abstract}

\maketitle

\section{Introduction}
Recent low-temperature transport experiments, utilizes the quantum Hall based interferometers to investigate the quantum nature of particles. Particularly, the Aharonov-Bohm (AB) interference experiments became a paradigm~\cite{Heiblum05:abinter,Goldman05:155313}, which infers the AB phases of both the electrons and the quasi- particles. The single particle edge-state picture is used to describe transport, meanwhile electrostatics is enriched by interactions and both are used to explain the observed AB oscillations~\cite{Goldman05:155313,Neder06:016804}. However, the actual spatial distribution of the edge-states is still under debate for real samples, since handling the full electrostatics is a formidable task, although, several powerful techniques are used~\cite{igor08:ab}. By full electrostatics we mean both handing the crystal growth parameters and the ``edge" definition of the interferometer, \emph{i.e.} gate, etched or trench-gated.

In this work, we provide a semi-analytical scheme to model AB interferometers induced on a two dimensional electron gas (2DEG) by solving the 3D Poisson for the given hetero-structure~\cite{Andreas03:potential}. Our calculation scheme also takes into account the lithographically defined surface patterns to obtain the electron and potential distributions under quantized Hall conditions~\cite{Sefa08:prb,Engin:09japon}. The distinct part of our calculation is that we can handle both gate and etching defined geometries. Our findings show that the etching defined samples provide a sharper potential profile than that of gate defined~\cite{Sefa08:prb}. In addition we can define the structure with trench gating, which is the case for the experiments, and show that the formation of the edge-states is strongly influenced.
\begin{figure}[h]
\centering
\includegraphics[scale=0.25]{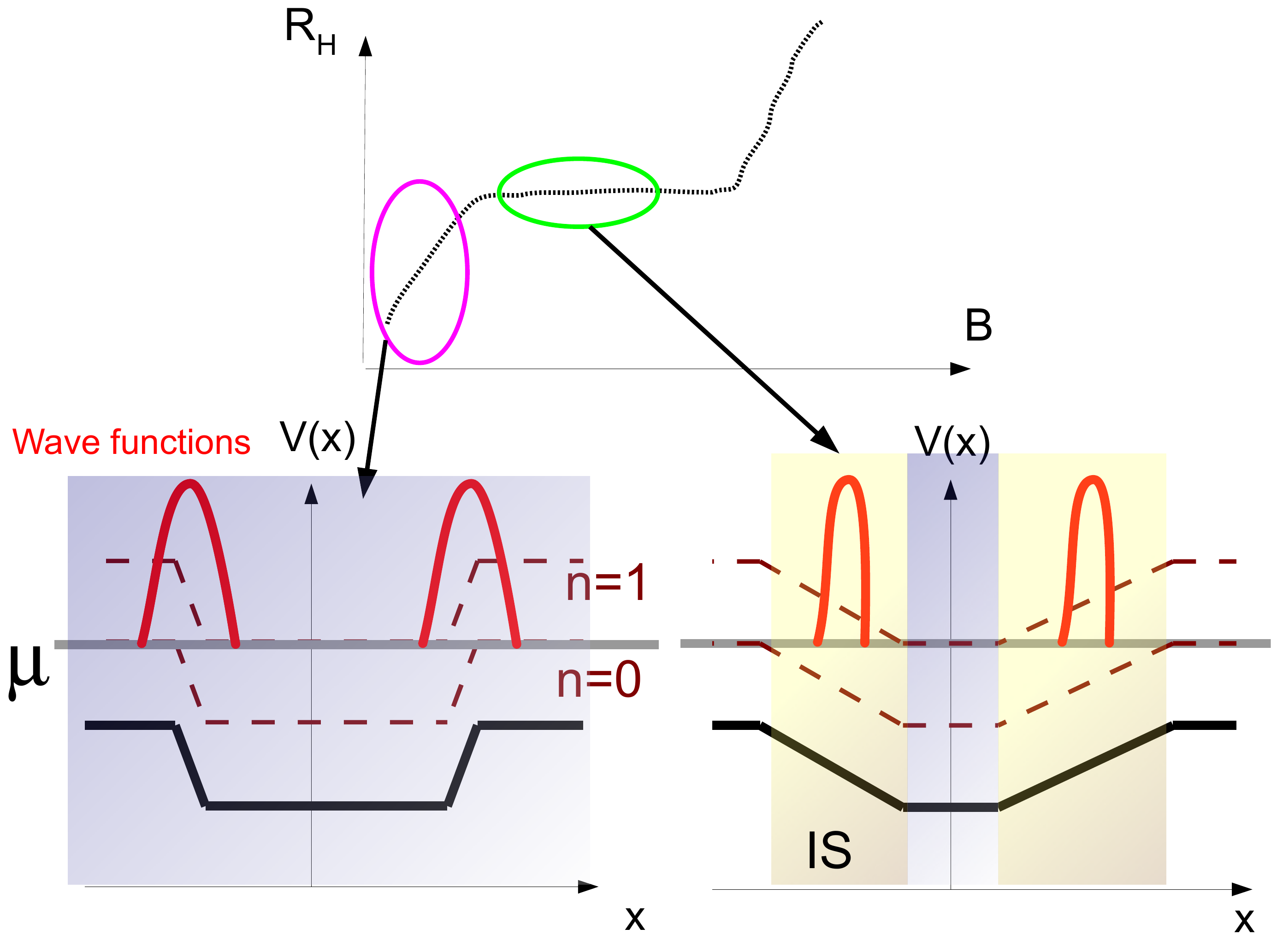}
\caption{Schematic presentation of the Hall resistance as a function of $B$ field (upper panel), together with the corresponding potential (thick black line), Landau levels (broken lines) and wave functions (red thick curves, lower panels), whereas $\mu$ denotes the chemical potential (or Fermi energy, at equilibrium and in 2D). The ellipsis indicate the $B$ field interval where ISs become leaky (left) or not (right).}\label{fig:fig1}
\end{figure}
\section{Interactions: Incompressible strips and Numerical results}
The high quality GaAs/AlGaAs hetero-structures provide a great laboratory for the condensed matter physicists to perform experiments inferring the phase of the particles and quasi-particles~\cite{Heiblum03:415,Heiblum05:abinter,Goldman05:155313}. Usually, an interferometer is defined on a ultra-high mobility ($\sim 10^6-10^7$ cm$^2$/V.s) wafer by means of etching and/or gating~\cite{Heiblum03:415,Roddaro05:156804,Roche07:QHP,Litvin07:033315} and so-called edge-states~\cite{Halperin82:2185,Buettiker86:1761,Chklovskii92:4026,siddiki2004} are utilized as phase-coherent ``beams'' to manipulate the interference pattern. These edge states are direct sequence of Landau quantization due to the perpendicular magnetic field and bending of them due to the finite size of the physical system. First considerations of the edge-states neglect the electron-electron interaction and base their models on 1D ballistic channels~\cite{Halperin82:2185,Buettiker86:1761} to explain the integer quantized Hall effect (IQHE). However, later interactions were also taken into account and the 1D edge states were replaced by compressible/incompressible strips~\cite{Chang90:871,Chklovskii92:4026,Lier94:7757,siddiki2004}. Essentially, Chklovskii \emph{et.al} attributed the properties of the 1D channels to the compressible strips where the Fermi energy (locally) equals the Landau energy. Hence, the current is carried by the compressible strips~\cite{Chklovskii92:4026}. In contrast, A.~M.~Chang and others claimed that the current is flowing from the incompressible strips due to the absence of back-scattering~\cite{Guven03:115327,siddiki2004}, since Fermi energy falls in between two consequent Landau levels both the electric field and conductivity vanish locally. All the above models provide a reasonable explanation to the IQHE, however, the 1D channel and compressible strip pictures both require bulk (localized) states to infer the transitions between the Hall plateaus. Meanwhile, the incompressible strip picture is almost self-standing and provides a self-consistent model both to describe electrostatics and transport.

Although, the incompressible picture sounds reasonable in explaining the IQHE unfortunately, it is a challenge to explain how to inject current from the contacts to these strips due to their ``incompressibility''~\cite{siddiki:epl}. Moreover in the case of interference, partitioning should take place between these incompressible strips which is apparently ``difficult''. Here, we would like to answer this question as quantitative as possible. First of all, in experiments the Aharonov-Bohm oscillations are observed in between the plateaus~\cite{Goldman05:155313,Neder06:016804,Bernd:ABosc.}. This means that the outermost edge channels are already much narrower than the magnetic length $l$ ($l^2=\hbar/eB$, where $\hbar$ is the Planck constant divided by two $\pi$, $e$ is the charge and $B$ represents the magnetic field strength), hence become ``leaky''. In the sense that the widths of the outermost incompressible strips are narrower than the quantum mechanical length scales. The models which consider many compressible strips utilize the Thomas-Fermi approximation that fail if the potential landscape vary strongly on the scale of the wave function (approximately the magnetic length) and this is exactly the case at the interference field interval. As an illustration we show the potential landscape at a certain cut across a Hall bar in Fig.~\ref{fig:fig1} (lower panels, black thick line), together with the Hall resistance (upper panel) and approximate wave functions (thick curves, lower panel). On one hand, once the incompressible strips become narrower than the wave extend, the strips are no longer incompressible as seen in the left panel of Fig.~\ref{fig:fig1}, which occurs at the lower side of the quantized Hall plateau. On the other hand, within the plateau a well developed incompressible strip (IS) exists, which decouples the Hall probe contacts. This makes us to conclude that, the partitioning can take place between the ``leaky'' ISs, which occurs only out of the plateau regime. The next question to be answered is ``why we do not observe interference in some samples ?''. The answer is quite easy: One should bring these strips to close enough vicinity, so that scattering can take place between two channels. Therefore, next we present some of our numerical results which treat a real interference device patterned on a 2DES by means of trench gating~\cite{Goldman05:155313}.
\begin{figure}[h]
\centering
\includegraphics[scale=0.25]{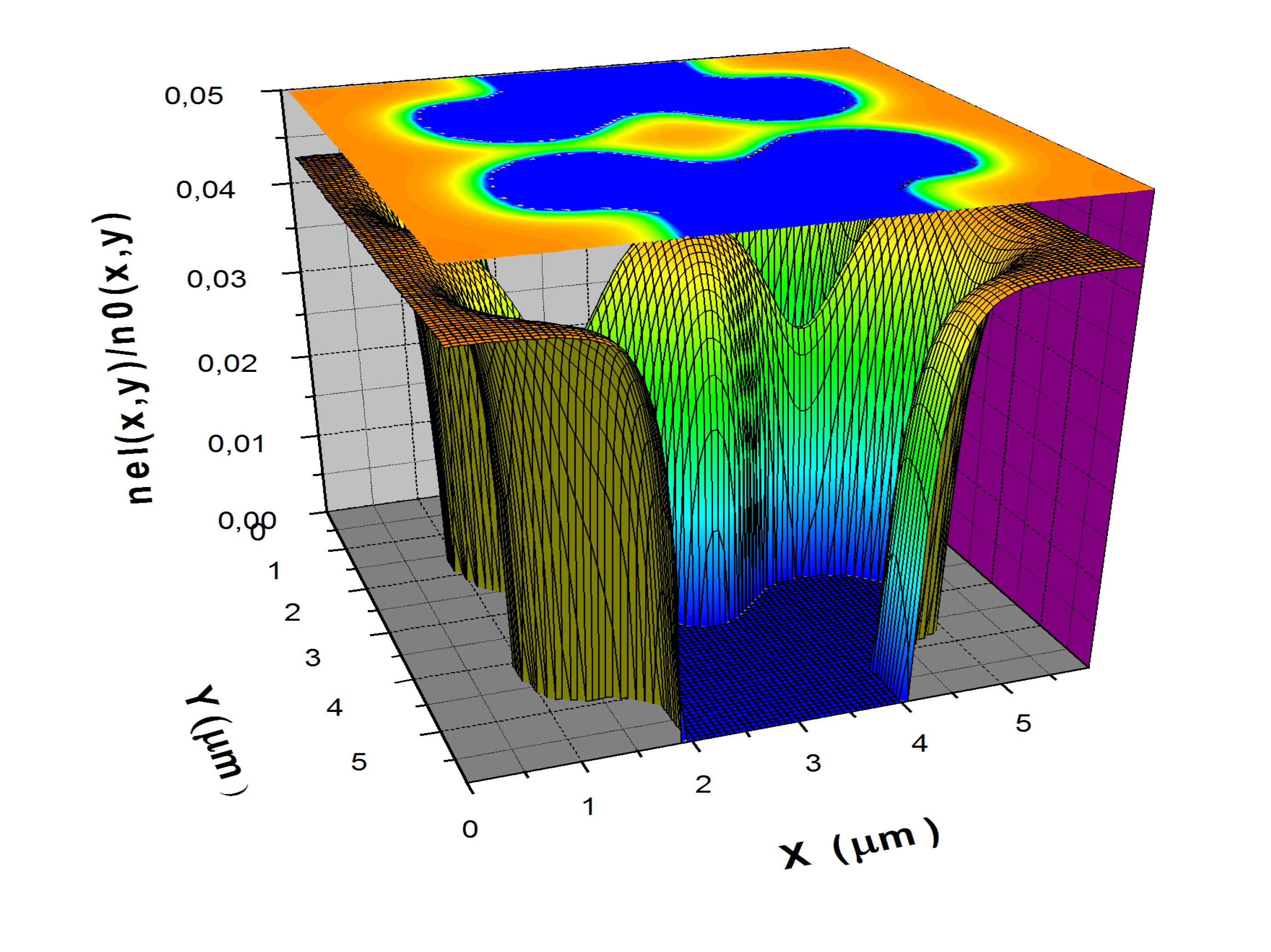}
\caption{The electron density as a function of lateral coordinates. Electrons are depleted at the barriers due to the trench gating, where -1.8 V is applied. Density is normalized with the dopant concentration to compare with the experiments.}\label{fig:fig2}
\end{figure}
We obtain the actual electron density distribution starting from the real crystal growth parameters, here we use the structural information provided by V. J. Goldman. The numerics is as follows: first the crystal parameters are given as an input which contains the dielectric constants of the material (12.4 for GaAs/AlGaAs structures) together with the dopant density (here, delta doped Silicon) and the location of the interface where 2DES forms. Next, we solve the Poisson equation considering open boundary conditions, whereas the surface pattern is defined by gates~\cite{Andreas03:potential}, chemical etching~\cite{Sefa08:prb} or a combination of these~\cite{Engin:09japon}. The electron density and potential distribution all over the structure is obtained self-consistently utilizing the 4$^{th}$ order grid technique and fast Fourier transformation. The iteration process is optimized not only by the clever choice of mesh points but also by successive over relaxation. Fig~\ref{fig:fig2}, depicts the electron density $n_{\rm el}(x,y)$ when the quantum dot is induced by trench gates under experimental conditions. The first test that our numerics has to go through is the estimation of the bulk and island electron density, reported experimentally~\cite{Goldman05:155313}. We, see that the central electron density is almost the same as reported in the experiment, \emph{i.e.} deviates $\% 5.6$ from the bulk density. Hence, the full electrostatic treatment of the interference device yields reliable results comparable with the experimental findings.

\begin{figure}[h]
\centering
\includegraphics[scale=0.25]{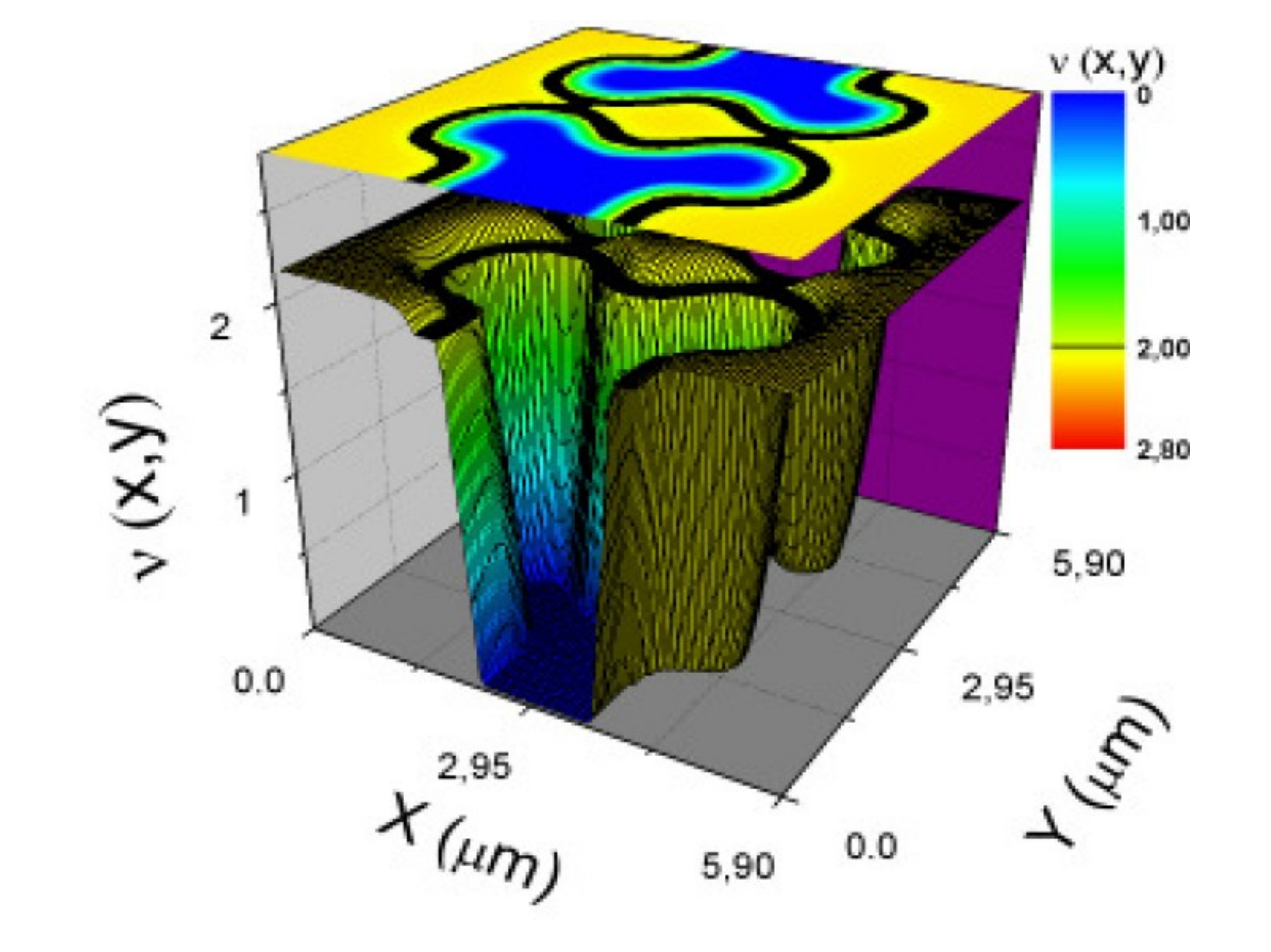}
\caption{The local filling factor as a function of lateral coordinates. Electron density is constant within the ISs, since there are no available states at the Fermi level and screening is poor. The two edge-states merge at the barriers, hence no partitioning can take place: No interference.}\label{fig:fig3}
\end{figure}

\begin{figure}[h]
\centering
\includegraphics[scale=0.25]{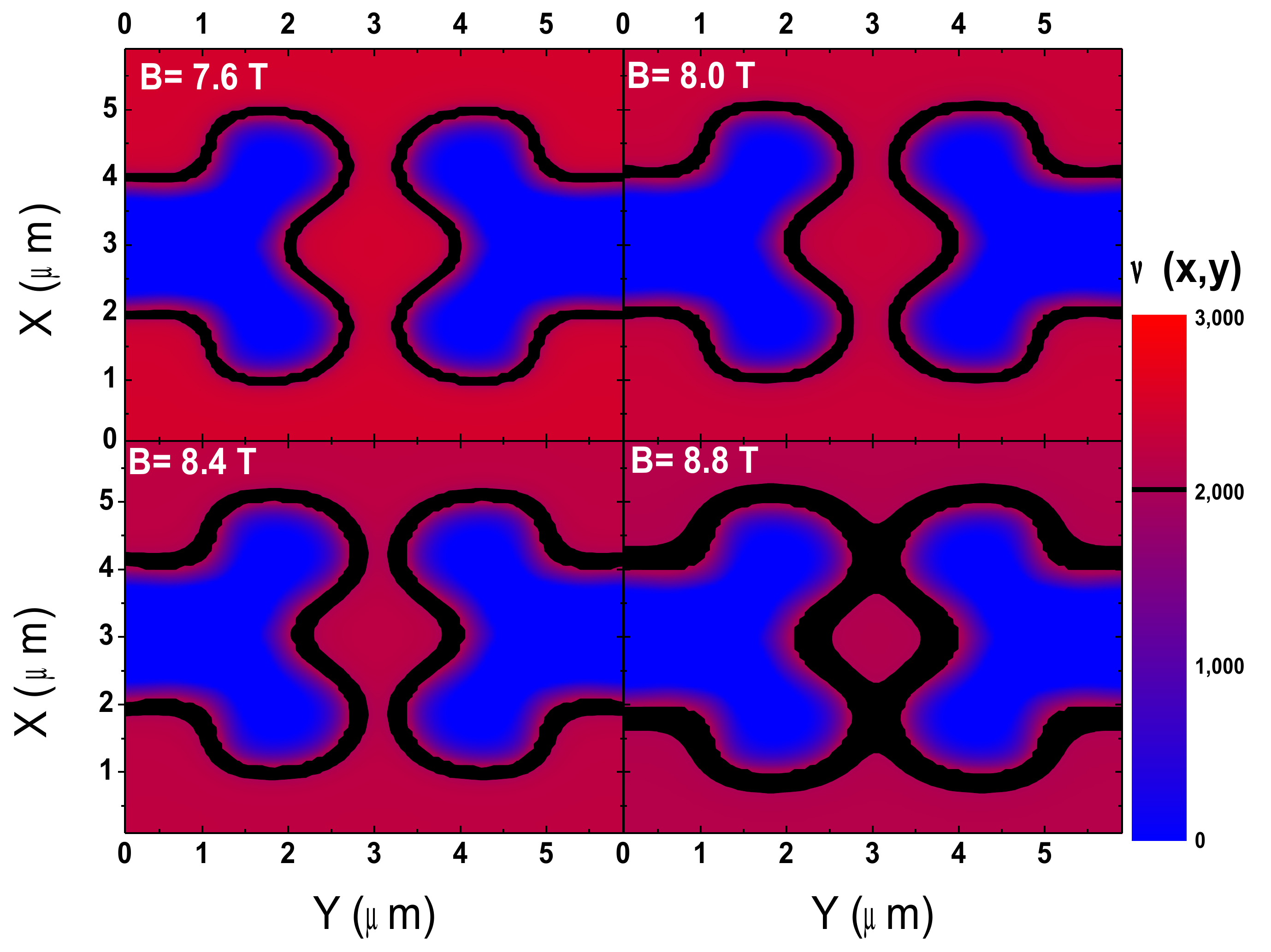}
\caption{Spatial distribution of ISs (black) calculated at four different field strengths at $T=1$ K. It is expected that only at $B=8.0$ T one can observe AB oscillations, the other cases prevent scattering between the edge-states.}\label{fig:fig4}
\end{figure}

Next, we calculate the spatial distribution of the ISs considering a strong perpendicular magnetic field. In principle, if one knows the electron distribution at $B=0$, it is somewhat easy to estimate the positions of the ISs. However, as we have discussed it is the widths of the incompressible strips that determines whether interference take place or not. Moreover, the analytical formulas given in Ref.~\cite{Chklovskii92:4026} to estimate the widths cannot be used here, since the potential landscape varies strongly on the scale of wave extend. Therefore, we also solve the Poisson and Schr\"odinger equations self-consistently in the presence of a strong $B$ field, a typical filling factor distribution is shown in Fig.~\ref{fig:fig3}, we note that the filling factor is nothing but is the electron density normalized by the field strength and is given by $\nu(x,y)=2\pi l^2n_{\rm el}(x,y)$. Note the similarity between the electron density and the filling factor, the most important difference is that now at the incompressible regions (black highlighted regions) the electron density is constant. To investigate interference conditions we also performed calculations at various magnetic field strengths of which we present in Fig.~\ref{fig:fig4} for selected values. At the highest $B$ shown here, the two incompressible strips overlap hence there is no scattering which means we are still in the plateau regime, $B=8.8$ T.  At a lower $B$ two well developed ISs come close to each other, however, since they are yet incompressible, as illustrated in Fig.~\ref{fig:fig1} lower right panel, one cannot observe interference: Scattering is strongly suppressed. The interference oscillations can be observed at $B=8.0$ T, since the incompressible strips are leaky and also are in close proximity, so that scattering can take place. The lowest magnetic field also prevents AB oscillations, since now the two channels are far appart from each other. This scenario is repeated for the next (even) integer filling factor plateau. Note that in our calculations we neglected the spin degree of freedom since the formation of the incompressible strips a is general situation given the single particle energy gap.

\section{Conclusions}
In this work we reported our numerical findings concerning simulations of real interference devices. The electron density and the spatial distribution of the incompressible strips are obtained self-consistently within a mean field approximation at the Hartree level. We showed that, the interference can take place only if the incompressible strips become leaky and come close to each other few magnetic lengths, so that scattering (hence partitioning) can take place. The actual calculation of the conductance and the AB oscillations is left untouched, however, a recent model is available to obtain these quantities which we currently work on~\cite{Bernd:ABosc.}. The scattering mechanisms and conductivities discussed here are merely phenomenological, therefore it is a great interest for us to perform calculations which also take these quantities into account in a more quantitative manner.

\ack
A.S. would like to thank to M. Heiblum, N. Offek, and L. Litvin for the experimental discussions, meanwhile we are also grateful to V.J. Goldman for providing us the sample structure and device pattern. F. Marquardt and B. Rosenow are appreciated for their fruitful and critical theoretical discussions. A partial financial support is given by TUBITAK under grant No.109T083 and by the Feza G\"ursey institute while organizing the 3$^rd$ nano-electronics symposium.

\section*{References}

\begin{thebibliography}{10}

\bibitem{Heiblum05:abinter}
M.~{Avinun-Kalish}, M.~{Heiblum}, O.~{Zarchin}, D.~{Mahalu}, and V.~{Umansky}.
\newblock {Crossover from `mesoscopic' to `universal' phase for electron
  transmission in quantum dots}.
\newblock {\em Nature}, 436:529--533, July 2005.

\bibitem{Goldman05:155313}
F.~E. Camino, W.~Zhou, and V.~J. Goldman.
\newblock Aharonov-bohm electron interferometer in the iqh regime.
\newblock {\em Phys. Rev B}, 72:155313, 2005.

\bibitem{Neder06:016804}
I.~Neder, M.~Heiblum, Y.~Levinson, D.~Mahalu, and V.~Umansky.
\newblock Unexpected behavior in a two-path electron interferometer.
\newblock {\em Phys. Rev. Lett.}, 96:016804, 2006.

\bibitem{igor08:ab}
S.~{Ihnatsenka} and I.~V. {Zozoulenko}.
\newblock {Interacting electrons in the Aharonov-Bohm interferometer}.
\newblock {\em ArXiv e-prints}, 803, March 2008.

\bibitem{Andreas03:potential}
A.~{Weichselbaum} and S.~E. {Ulloa}.
\newblock {Potential landscapes and induced charges near metallic islands in
  three dimensions}.
\newblock {\em Phys. Rev. E}, 68(5):056707--+, November 2003.

\bibitem{Sefa08:prb}
S.~{Arslan}, E.~{Cicek}, D.~{Eksi}, S.~{Aktas}, A.~{Weichselbaum}, and
  A.~{Siddiki}.
\newblock {Modeling of quantum point contacts in high magnetic fields and with
  current bias outside the linear response regime}.
\newblock {\em Phys. Rev. B}, 78(12):125423--+, September 2008.

\bibitem{Engin:09japon}
E.~{Cicek}, A.~I. {Mese}, M.~{Ulas}, and A.~{Siddiki}.
\newblock {Spatial Distribution of the Incompressible Strips at Aharonov-Bohm
  Interferometer}.
\newblock {\em ArXiv e-prints}, September 2009.

\bibitem{Heiblum03:415}
Y.~Ji, Y.~Chung, D.~Sprinzak, M.~Heiblum, D.~Mahalu, and H.~Shtrikman.
\newblock {\em Nature}, 422:415, 2003.

\bibitem{Roddaro05:156804}
S.~{Roddaro}, V.~{Pellegrini}, F.~{Beltram}, L.~N. {Pfeiffer}, and K.~W.
  {West}.
\newblock {Particle-Hole Symmetric Luttinger Liquids in a Quantum Hall
  Circuit}.
\newblock {\em Physical Review Letters}, 95(15):156804--+, October 2005.

\bibitem{Roche07:QHP}
P.~{Roulleau}, F.~{Portier}, D.~C. {Glattli}, P.~{Roche}, A.~{Cavanna},
  G.~{Faini}, U.~{Gennser}, and D.~{Mailly}.
\newblock {Direct measurement of the coherence length of edge states in the
  Integer Quantum Hall Regime}.
\newblock {\em ArXiv e-prints}, 710, October 2007.

\bibitem{Litvin07:033315}
L.~V. {Litvin}, H.-P. {Tranitz}, W.~{Wegscheider}, and C.~{Strunk}.
\newblock {Decoherence and single electron charging in an electronic
  Mach-Zehnder interferometer}.
\newblock {\em Phys. Rev. B}, 75(3):033315--+, January 2007.

\bibitem{Halperin82:2185}
B.~I. Halperin.
\newblock {\em Phys. Rev. B}, 25:2185, 1982.

\bibitem{Buettiker86:1761}
M.~B{\"u}ttiker.
\newblock Four-terminal phase-coherent conductance.
\newblock {\em Phys. Rev. Lett.}, 57:1761, 1986.

\bibitem{Chklovskii92:4026}
D.~B. Chklovskii, B.~I. Shklovskii, and L.~I. Glazman.
\newblock Electrostatics of edge states.
\newblock {\em Phys. Rev. B}, 46:4026, 1992.

\bibitem{siddiki2004}
A.~Siddiki and R.~R. Gerhardts.
\newblock Incompressible strips in dissipative hall bars as origin of quantized
  hall plateaus.
\newblock {\em Phys. Rev. B}, 70:195335, 2004.

\bibitem{Chang90:871}
A.~M. Chang.
\newblock A unified transport theory for the integral and fractional quantum
  hall effects: phase boundaries, edge currents, and transmission/reflection
  probabilities.
\newblock {\em Solid State Commun.}, 74:871, 1990.

\bibitem{Lier94:7757}
K.~Lier and R.~R. Gerhardts.
\newblock Self-consistent calculation of edge channels in laterally confined
  two-dimensional electron systems.
\newblock {\em Phys. Rev. B}, 50:7757, 1994.

\bibitem{Guven03:115327}
K.~G{\"u}ven and R.~R. Gerhardts.
\newblock Self-consistent local-equilibrium model for density profile and
  distribution of dissipative currents in a hall bar under strong magnetic
  fields.
\newblock {\em Phys. Rev. B}, 67:115327, 2003.

\bibitem{siddiki:epl}
A.~{Siddiki}.
\newblock {Current direction induced rectification effect on (integer)
  quantized Hall plateaus}.
\newblock {\em ArXiv e-prints:0805.2312[cond-mat.mes-hall]}, May 2008.

\bibitem{Bernd:ABosc.}
D.~T. {McClure}, Y.~{Zhang}, B.~{Rosenow}, E.~M. {Levenson-Falk}, C.~M.
  {Marcus}, L.~N. {Pfeiffer}, and K.~W. {West}.
\newblock {Edge-State Velocity and Coherence in a Quantum Hall Fabry-P{\'e}rot
  Interferometer}.
\newblock {\em Physical Review Letters}, 103(20):206806--+, November 2009.

\end{thebibliography}
\bibliographystyle{unsrt}

\end{document}